\documentclass[fleqn,usenatbib]{mnras}

\usepackage{newtxtext,newtxmath}
\usepackage[T1]{fontenc}

\DeclareRobustCommand{\VAN}[3]{#2}
\let\VANthebibliography\thebibliography
\def\thebibliography{\DeclareRobustCommand{\VAN}[3]{##3}\VANthebibliography}

\usepackage{graphicx}	
\usepackage{amsmath}	
\usepackage{CJKutf8}

\title[SMT Black Hole Spins]{A Diverse Distribution of Black Hole Spins from Stable Mass Transfer}

\author[Ma et al.]{
Linhao Ma \begin{CJK*}{UTF8}{gbsn}(马林昊)\end{CJK*},$^{1,2}$\thanks{E-mail: linhaoma@princeton.edu}
Jakub Klencki,$^{3}$
Eliot Quataert,$^{1}$
and Lieke van Son$^{1,4,5}$
\\
$^{1}$Department of Astrophysical Sciences, Princeton University, 4 Ivy Lane, Princeton, NJ 08544, USA\\
$^{2}$Kavli Institute for Theoretical Physics, University of California, Santa Barbara, CA 93106, USA\\
$^{3}$Max Planck Institute for Astrophysics, Karl-Schwarzschild-Strasse 1, 85748 Garching, Germany\\
$^{4}$Department of Astrophysics, Radboud University, PO Box 9010, NL-6500 GL Nijmegen, The Netherlands\\
$^{5}$Center for Computational Astrophysics, Flatiron Institute, 162 Fifth Avenue, New York, NY 10010, USA
}

\date{Accepted XXX. Received YYY; in original form ZZZ}

\pubyear{\the\year{}}

\begin{document}
\label{firstpage}
\pagerange{\pageref{firstpage}--\pageref{lastpage}}
\maketitle

\begin{abstract}
Gravitational wave observations have found over 300 merging binary black holes, yet their origins remain uncertain. Recent work showed that many may come from isolated stellar binaries whose orbits shrink through stable mass transfer. If true, their spins may help to distinguish this channel from other formation pathways. We investigate the tidal spin up of black hole progenitor stars with detailed modeling of binaries undergoing stable mass transfer. We calculate the tidal torques by solving tidally excited oscillation modes and predict the resulting black hole spins. We find a diverse spin distribution strongly affected by the mass transfer histories of the progenitors. Binaries can form black holes with moderate spins ($0.1\lesssim\chi_\mathrm{eff}\lesssim0.3$) if they only go through case A or case B mass transfer. In the former case, they can become super-synchronized upon detachment, while in the latter case, the donor is usually only partially stripped, leaving a puffy envelope where strong tides are excited. If both case A and case AB mass transfer occur, the resulting black hole spins are almost always negligible. As the mass transfer history is jointly determined by mass ratio and initial binary period, our results predict an anti-correlation between black hole spins and mass ratio, consistent with limited evidence from data. Our results can also potentially explain the case of GW190412, a moderately-spinning binary with a high mass ratio. We discuss the limitations of our methods and additional physics (e.g., nonlinear tides, case C, and L2 mass transfer) that need to be incorporated in future work.
\end{abstract}




\section{Introduction}

Since the first detection of a gravitational wave event in 2015, over three hundred merging binary black holes (BBHs) have been observed \citep{GWTC-4.0.1,GWTC5.0.1,GWTC-5.0,LIGO2025}, yet we still do not have a clear understanding of how these systems formed. Different formation channels have been proposed that, in principle, produce different BBH populations with unique observational signals capable of distinguishing them. However, the exact theoretical predictions rely on detailed modeling of complicated physical processes that remain a major difficulty for most channels.

It has been proposed for decades that isolated binary evolution is one of the ways to form merging binary black holes (e.g., \citealt{Mapelli2021,Mandel2022b,Mandel2022}). Such binaries are typically born with two massive stars in orbits much wider than those of merging compact objects. An important question is how their orbits eventually tighten to form binary black holes that can merge within a Hubble time. It was initially believed that common envelope (CE) ejection is responsible for this process due to its high efficiency in shrinking the binary orbits. Therefore, CE ejection has been widely assumed in rapid binary population synthesis models (see, e.g., \citealt{Dominik2012,Mennekens2014,Belczynski2016,Eldridge2016,Mapelli2018,Kruckow2018,Breivik2020}).

In recent years, however, detailed binary modeling with a more sophisticated treatment of binary mass exchange has challenged this picture. Several authors have shown that CE ejections are often not successful, and massive binaries are more likely to form merging BBHs after one or more phases of stable mass transfer (SMT) that tighten their orbits \citep{Inayoshi2017,vandenHeuvel2017,Klencki2021,Marchant2021,Gallegos-Garcia2021,vanSon2022,Picco2024,Xu2025}. One of the observables that may help distinguish between formation channels is the spin of merging BBHs. For example, by combining binary population synthesis with prescriptions for tidal interactions, \cite{Bavera2021b} found that systems with non-negligible spins form only through the CE ejection channel, whereas stable mass transfer produces only low-spin BBHs.

In the isolated binary evolution channel, BBHs usually get their spin from a prior tidal spin up phase of their progenitors. Otherwise, the strong core-envelope coupling leads to low spins as stars lose their envelope before core collapse \citep{FullerMa2019}. Previous theoretical works often focus on this phase, assuming the idealized case of naked helium stars with an already formed black hole companion (e.g., \citealt{Bavera2020,Bavera2021,Ma2023}), which is a natural outcome of common envelope ejection. However, stable mass transfer may change this picture, as both the structure and evolution history of the stripped star become more complicated.

In this work, we calculate the tidal spin-up of black hole progenitor stars that went through stable mass transfer to shrink their binary orbits. We compute detailed binary evolutionary models with self-consistent treatments of envelope stripping that produce realistic stellar structures after the mass transfer phase. We improve upon previous works that largely include tidal physics as a parameterization (e.g., \citealt{Qin2018,Bavera2021,Bavera2021b}), by calculating the tidal excitation of self-damping oscillation modes in detailed binary evolutionary models that treat mass transfer consistently. This physically motivated treatment is currently the most accurate method that can be applied to stellar evolutionary models, and it allows us to make the most accurate predictions on the spins of BBHs formed through the stable mass transfer channel to date.

We find that stable mass transfer can indeed lead to a diverse distribution of black hole spins that depends on the mass transfer history of the binary. Different histories can lead to fast-spinning, partially stripped helium stars and super-synchronized binaries that are not expected from the conventional treatments of tidal spin up for BBH progenitor stars. As the evolutionary history correlates with other binary parameters, such as the mass ratio of the binary, it creates an additional correlation between BBH spins and other observables.

The structure of this paper is as follows: in Section \ref{sec:methods} we describe the methods we use, including binary models and tidal spin up calculations. In Section \ref{sec:results} we show our results on how different mass transfer histories can lead to different BBH spins, and summarize the possible pathways in Section \ref{sec:results-pathways} and Figure \ref{fig:flowchart}. We discuss the implications of our results, including a possible anti-correlation between BBH spins and mass ratio, an explanation for GW190412, and a comparison to previous works on helium-star-black-hole binaries in Section \ref{sec:discussions}, where we also discuss the limitations and uncertainties of our work. We conclude in Section \ref{sec:conclusions}.

\section{Methods}
\label{sec:methods}

Here we describe our model's setup and our methods for calculating tidal torques and black hole spins.

\subsection{Stellar Models}
\label{sec:models_setup}

We ran a set of binary evolutionary models with the MESA stellar evolution code, r15140 \citep{Paxton2011,Paxton2013,Paxton2015,Paxton2018,Paxton2019}. Each model consists of a star  and a point mass in a circular orbit, which represents a donor star and a black hole that has already formed in the binary. We chose two different initial donor star masses, with $M_*=30\,M_\odot$ or $M_*=50\,M_\odot$. For each stellar mass, we run a grid of models, varying the initial binary mass ratio (defined as $q_\mathrm{i}=M_\mathrm{BH}/M_*$) from 0.14 to 0.5 in 0.01 intervals and the logarithm of the initial orbital period in days $\log(P_\mathrm{orb,i})$ from 0.2 to 3.6 in 0.04 intervals. Although these are not comprehensive grids for all BBH progenitors, the resulting second-born black hole masses, derived from the final helium core mass of the donor, range from 8 to 25 solar-masses, which are representative of most GW events (see, e.g., \citealt{GWTC-5.0} for the GWTC BH mass distribution).

The model setup we are using is similar to \cite{Klencki2025,Klencki2025b}, and we briefly summarize it below. We run our binary models from the pre-main sequence until the donor depletes its core carbon (defined as the central mass fraction of carbon dropping below 0.1\%), shortly before core collapse. We chose a universal 10\% solar-metallicity ($Z=0.0017$, \citealt{Grevesse1996}) for all our models. We implemented the same controls for convection, semi-convection, mixing, stellar winds, and timesteps as in \cite{Klencki2025}. 

Unlike \cite{Klencki2025}, we turned off stellar rotation, and the binary orbits in our models evolve only due to mass transfer and mass loss. The reason is that the default tidal module in MESA, based on \cite{Hurley2002}, is not suitable for hot stars and can overestimate the tidal torques acting on helium stars by several orders of magnitude. Our own method, however, can only be applied by post-processing non-rotating stellar evolutionary models (see Section \ref{sec:methods_tidal_torques}). This naturally causes some inconsistencies, as many physical processes depend on stellar rotation. We discuss some effects in Section \ref{sec:caveats} and will look into them in future works.

We treat mass transfer in our stellar models with the prescription developed by \cite{Marchant2021}, which traces back to the method described in \cite{Kolb1990}. The mass transfer from the donor is treated consistently with stellar expansion and orbital shrinking, and the accretion to the BH is Eddington-limited. The non-accreted flow leaves the system with the specific angular momentum near the vicinity of the accretor \citep{Soberman1997}. On Roche-lobe overflow, mass loss through the outer Lagrangian point of the donor ($L_\mathrm{out}$, on the opposite side of $L_1$ relative to the donor) is also calculated self-consistently and carries the specific angular momentum at $L_\mathrm{out}$. Both of them take away angular momentum from the orbit and alter the orbital periods. When mass transfer exceeds a threshold rate of $1\,M_\odot\,\mathrm{yr}^{-1}$, it is considered unstable, and the binary model enters the CE phase.

We find that aside from a handful of models that do not numerically converge, the binaries in our grids can be classified into three types based on their evolutionary pathways and outcomes: i. binaries that undergo a common envelope phase; ii. binaries that avoid CE with a gravitational wave decay timescale \citep{Peters1964}
\begin{equation}
t_\mathrm{GW\,decay}=\frac{5}{256}\frac{a^4}{(G^3/c^5)M_*M_\mathrm{BH}(M_*+M_\mathrm{BH})}<13.8\,\mathrm{Gyr}\,
\end{equation}
at the end of their evolution, where $a$ is the binary separation; and iii. binaries with $t_\mathrm{GW\,decay}\geq 13.8\,\mathrm{Gyr}$. In this work, we are interested in calculating the tidal evolution of the second type of binaries that can merge within a Hubble time.

\subsection{Tidal Torque Calculation}
\label{sec:methods_tidal_torques}

We calculate the tidal torque on the donor star by the dissipation of internal gravity waves excited by the tidal potential of their companion. This is known as ``dynamical tides'', which is different from the conventional picture of ``equilibrium tides'', whose turbulent viscous dissipation is not significant in massive hot stars with small or vanishing surface convective layers \citep{Zahn1975,Zahn1977}. We mainly consider linear radiative damping of these waves, but we discuss nonlinear effects in Section \ref{sec:nonlinear}.

We adopt two ways to calculate the tidal torques. The first is known as the ``mode decomposition'' method as described in \cite{Ma2023,Ma2024}. Specifically, the total torque on the donor star is given by the sum of torques from individually excited g-modes:
\begin{equation}
\label{eq:torque_mode}
\tau_\mathrm{tide}= -\sum_{\mathrm{mode}\,\alpha}\frac{2\omega_\alpha \gamma_\alpha q^2M_*R_*^2|W_{22}Q_\alpha|^2\omega_\mathrm{f}^2}{(\omega_\alpha-\omega_\mathrm{f})^2+\gamma_\alpha^2}\bigg(\frac{R_*}{a}\bigg)^6\,,
\end{equation}
where $R_*$ is the radius of the donor, and the other variables are defined in the same way as in \cite{Ma2024}, Equation 1. This parameter-free method is based on first-principles solutions of non-adiabatic oscillation modes for a given stellar structure, and it is the most accurate way to calculate dynamical tides in evolutionary models to date (see Section \ref{sec:caveats} for remaining uncertainties).

However, before mass transfer occurs, the binary can have a very wide orbit with a long orbital period. In this case, tidally excited g-modes are of extremely high-order, making them heavily damped and difficult to solve numerically. Nevertheless, these modes naturally enter ``the traveling wave limit'' originally described by \cite{Zahn1975}, where the torques can be approximated by the following fitting formula given by \cite{Kushnir2017}:
\begin{equation}
\label{eq:zahn_torque}
\tau_\mathrm{Zahn}=\beta_2\frac{GM_\mathrm{BH}^2}{r_\mathrm{c}}\bigg(\frac{r_\mathrm{c}}{a}\bigg)^6s_\mathrm{c}^{8/3}\frac{\rho_\mathrm{c}}{\bar{\rho}_\mathrm{c}}\bigg(1-\frac{\rho_\mathrm{c}}{\bar{\rho}_\mathrm{c}}\bigg)^2\,,
\end{equation}
where the definitions of variables are the same as in \cite{Ma2024}, Equation 8.

As we will see in Section \ref{sec:mass_transfer_history}, for systems that go through case B mass transfer, the tidal torque is only important after the rapid case B mass transfer phase because this phase removes the majority of the stellar angular momentum accumulated by previous processes of tidal spin-up. Therefore, for these systems, we use Equation \ref{eq:zahn_torque} to calculate the torque until the moment when the case B mass-transfer rate reaches its peak, after which we switch to Equation \ref{eq:torque_mode}. To test the validity, we setup additional runs with Equation \ref{eq:zahn_torque} scaled by a factor of 0.1 to 5.0 and we find the resulting black hole spins are unchanged.

\subsection{Tidal Spin-up Integration}
\label{sec:methods_tidal_spin_up}

With the expressions for the tidal torques, we can calculate the tidal spin-up of the donor stars by post-processing the binary models. We integrate the following equation:
\begin{equation}
    \Delta J_\mathrm{spin}=(\,\mathrm{sign}(\Omega_\mathrm{orb}-\Omega_\mathrm{spin})\tau_\mathrm{tide}+\tau_\mathrm{mass\,loss}\,)\,\Delta t\,,
\end{equation}
where $\tau_\mathrm{mass\,loss}=\dot{M_*}R_*^2\Omega_\mathrm{spin}$ is the rate of angular momentum loss due to mass loss, either due to stellar wind or mass transfer. The function $\mathrm{sign}(\Omega_\mathrm{orb}-\Omega_\mathrm{spin})$ approximately accounts for the negative tidal torques when the star is super-synchronized\footnote{Equations \ref{eq:torque_mode} and \ref{eq:zahn_torque} neglect the rotational effects on modes, which should be taken into account for super-synchronized torques when the star is fast rotating. Nevertheless, \cite{Fuller2014} shows that inverting the torque direction gives a good approximation in this case as long as the star is not close to synchronization.}.

We used first-order Eulerian integration, starting at 10,000 years after ZAMS, to avoid the initial thermal-relaxation phase of the donor star. We used interpolated values of stellar structures and oscillation mode solutions to calculate $\tau_\mathrm{tide}$ and $\tau_\mathrm{mass\,loss}$ at each time step, as described in detail in \cite{Ma2023}. For simplicity, we assumed rigid rotation to obtain $\Omega_\mathrm{spin}$ from $J_\mathrm{spin}$ with the stellar moment of inertia. This implicitly assumes efficient internal angular momentum transport in stars. Under this assumption, the initial rotation of the stellar core is not important because the majority of its angular momentum will be transported to the envelope that is eventually lost before core collapse \citep{MaFuller2019}. Therefore, we start integration with an initial $J_\mathrm{spin}=0$. We chose an adapted timestep control as described in Appendix \ref{app:integration_timesteps} for numerical convergence, and we tested that our results converge when the timestep is cut by half.

We note that, in principle, an inverse tidal torque acts on the orbit and changes $\Omega_\mathrm{orb}$. However, this effect is usually negligible as the moment of inertia of the orbit is much larger than the moment of inertia of the donor star. Hence $\Delta\Omega_\mathrm{orb}\sim(I_\mathrm{spin}/I_\mathrm{orb})\Delta\Omega_\mathrm{spin}\ll\Omega_\mathrm{orb}$. We therefore ignore this effect and take the orbital frequency from the binary models as if there were no tides acting on the orbit. Resolving this inconsistency requires a real-time calculation of tides on the binary evolution model and is beyond the capabilities of our current method.

We stop integration at the end of the donor's helium-burning phase, defined as the moment when the central mass fraction of $^4\mathrm{He}$ drops below 0.1\%. For simplicity, we assume that the helium core of the donor at this point will directly collapse into a black hole that conserves its angular momentum (see, e.g., \citealt{Boccioli2024} for more detailed discussions on this assumption). Hence, the final black hole spin is given by:
\begin{equation}
\label{eq:spin}
    a = \frac{cJ_\mathrm{He\,core}}{GM^2_\mathrm{He\,core}}\,.
\end{equation}

In our models, the star still has a few tens of thousands of years left before core-collapse. However, many of the assumptions we make above may break down beyond the helium-burning phase, and modeling for late-stage stellar evolution is very uncertain. In Section \ref{sec:discussions_case_C}, we discuss a possible case C mass-transfer process at this stage and how it may affect our results.

\section{Results}
\label{sec:results}

\begin{figure*}
    \centering
    \includegraphics[width=\textwidth]{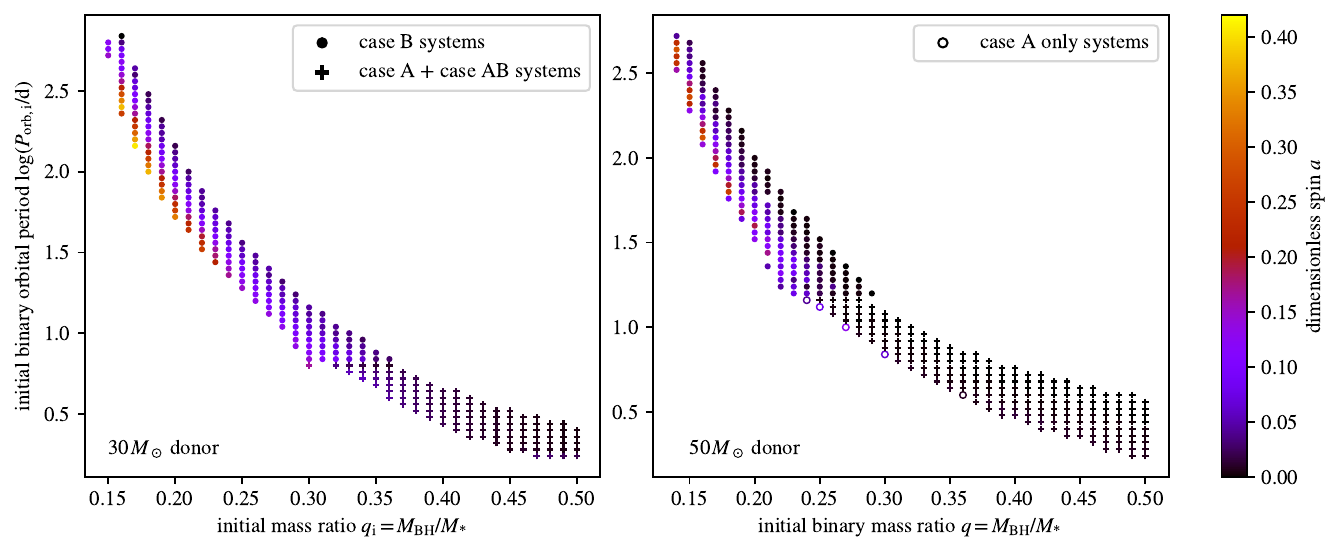}
    \caption{Results of black hole spin calculations in the $q_\mathrm{i}-\log(P_\mathrm{orb,i})$ space, for binaries that form through stable mass transfer and merge in a Hubble time. The spins are derived from Equation \ref{eq:spin} at core helium depletion. We see that binaries with a $30\,M_\odot$ donor can only go through case B mass transfer (filled circles), or case A with a following case AB mass transfer  phase (plus signs). The donor in the former case can become black holes with moderate ($a$ up to 0.4) spins, while in the latter case the spin becomes negligible ($a<0.1$). For binaries with a $50\,M_\odot$ donor, the results are similar but a few systems can experience case A mass transfer only (empty circles), with a small but non-negligible spins ($0.1\lesssim a\lesssim0.2$).}
    \label{fig:a_q_logPi}
\end{figure*}

\subsection{General Predictions for Black Hole Spin}
\label{sec:results-spins}

Figure \ref{fig:a_q_logPi} shows the results of our calculation for the binary model grids with donors of $30\,M_\odot$ and $50\,M_\odot$, respectively. We only show the systems that went through stable mass transfer and merge in Hubble time. We can see that these binaries form a ``banana shape'' in the space of $q_\mathrm{i}-\log(P_\mathrm{orb,i})$, with larger/smaller $\log(P_\mathrm{orb,i})$ as $q_\mathrm{i}$ becomes smaller/larger. This is because they are restricted to a relatively narrow range of separations, given by $a^3 = GM_*(1+q)P_\mathrm{orb}^2$. If the final separation is too large, the binaries will not merge in Hubble time and form gravitational wave sources; if the separation is too small, the mass transfer rate would become unstable, and the binaries would enter the common envelope phase. Our findings confirm the results shown by \cite{Marchant2021,Gallegos-Garcia2021,Klencki2025}.

As the donor evolves from ZAMS to core-helium depletion, we can classify these binaries into three types according to the history of their mass transfer processes, as shown in Figure \ref{fig:a_q_logPi}:
\begin{enumerate}
    \item \textbf{``Case B'' systems} (filled circles): binaries that only went through case B mass transfer phases \citep{Eggleton2006}, which occur after the donor evolves off the main sequence;
    \item \textbf{``Case A + Case AB'' systems} (plus signs): binaries that went through case A mass transfer phases on the main sequence \citep{Eggleton2006} before they went through at least one other case AB mass transfer phase after the main sequence;
    \item \textbf{``Case A only'' systems} (empty circles): binaries that only went through case A mass transfer phases on the main-sequence, without a later case AB mass transfer phase.
\end{enumerate}

The first two cases are separated by a roughly fixed initial orbital period. This is because the donor has to expand to a Roche lobe radius, given by the approximate formula from \cite{Eggleton1983}:
\begin{equation}
\begin{split}
    r_\mathrm{RL} &\approx \frac{0.49q^{-2/3}}{0.6q^{-2/3}+\ln(1+q^{-1/3})}a\\
    &=\frac{0.49q^{-2/3}(1+q)^{1/3}}{0.6q^{-2/3}+\ln(1+q^{-1/3})}(GM_*)^{1/3}P_\mathrm{orb}^{2/3}
\end{split}
\end{equation}
on the main-sequence to start case A mass transfer, which favors small $P_\mathrm{orb}$ and is relatively insensitive to $q$. The third case only occurs in a few models with a $50\,M_\odot$ donor, and they are at the lower edge of the parameter space with high mass transfer rates.

We show our black hole spin calculations with colors in Figure \ref{fig:a_q_logPi}. We see that different  mass transfer histories lead to different spin distributions. ``Case B'' binaries can have moderate dimensionless spins, up to $a\sim0.45$ for binaries with $30\,M_\odot$ donors and $a\sim0.35$ for those with $50\,M_\odot$ donors. The few ``case A only'' systems can also get non-negligible spins, typically up to $a\sim0.2$. On the other hand, the spins for the ``case A + case AB'' systems are almost always low, not exceeding $a\sim0.1$. In all cases the spins tend to get higher as the initial binary orbital periods get shorter, which is expected as tides become stronger with tighter orbits.

\subsection{Effect of Different Mass Transfer Histories}
\label{sec:mass_transfer_history}

\begin{figure}
    \centering
    \includegraphics[width=\columnwidth]{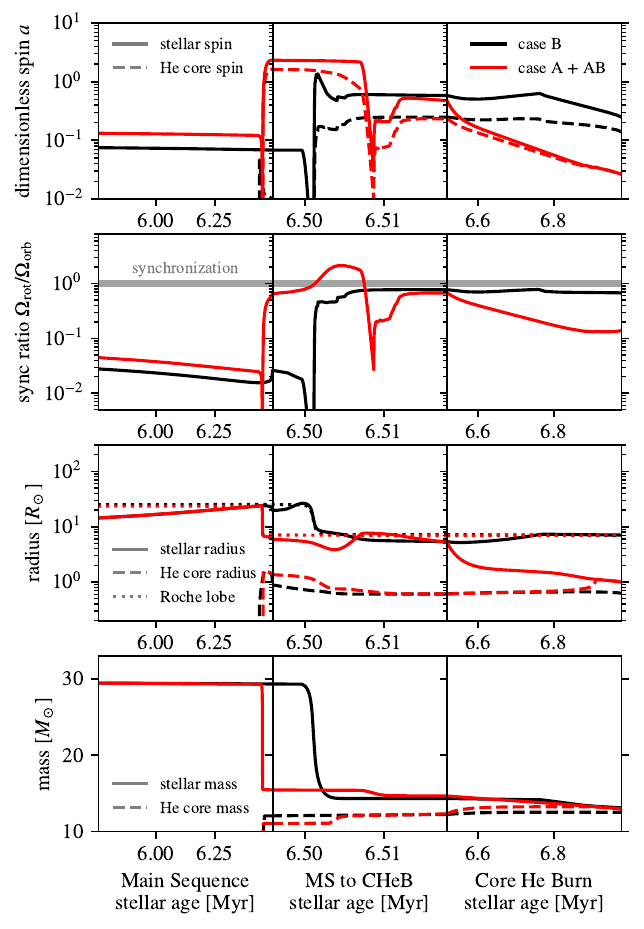}
    \caption{Comparison of the tidal spin-up history between one model that only went through case B mass transfer (black lines) and another that went through case A + AB mass transfer (red lines). The left, middle, and right sections show different quantities zoomed in during the later stage of main-sequence, between main-sequence and core-helium ignition, and during the core-helium burning phase, respectively. We see that while the donors lose similar amount of masses at the end, the ``case B'' system has higher spin than the ``case A + AB'' system. This is because while they both lose the majority of their angular momenta during the rapid case B/AB mass transfer episode (occurring at  6.50 and 6.51 Myrs respectively), the ``case B'' donor ends up as a much puffier star ($R_*\sim 7\,R_\odot$) with a significant hydrogen envelope over its helium core, feeling a much stronger tidal torque during the helium burning phase.}
    \label{fig:compare_AB_B}
\end{figure}

To understand why different mass transfer histories can lead to different spin distributions, we take a closer look at the evolution and tidal spin-up history of a few representative models.

Figure \ref{fig:compare_AB_B} shows the comparison between a ``case B'' binary (black lines; $M_*=30M_\odot$, $q_\mathrm{i}=0.33$, and $\log(P_\mathrm{orb,i})=0.84$) and a ``case A + case AB'' binary (red lines; $M_*=30M_\odot$, $q_\mathrm{i}=0.33$, and $\log(P_\mathrm{orb,i})=0.80$). These two models are close to each other in the $q_\mathrm{i}-\log(P_\mathrm{orb,i})$ space, so they evolve similarly before mass transfer occurs. In the upper and lower panels, we show the spin parameters and masses of the donor star and its helium core, defined as $a=GJ_*/(c^2M_*)$ and $a_\mathrm{He}=GJ_\mathrm{He\,core}/(c^2M_\mathrm{He\,core})$. We can see that while the helium core masses are similar throughout the evolution, the ``case B'' model produces a much higher spin magnitude than the ``case A + case AB'' model, confirming our findings in Section \ref{sec:results-spins}.

As the mass of the donor does not change by orders of magnitude as it evolves, the spin parameter roughly traces its total angular momentum, which is accumulated by tides and lost through mass loss. We see that while the star can initially get some spin from tides, almost all its angular momentum is removed during the rapid case B/AB mass transfer phase, occurring between 6.50 and 6.51 Myrs in these two models. The final black hole can gain its angular momentum only from tidal interactions during the core-helium-burning phase of its progenitor.

However, while the two models have similar initial orbital periods with similar companions (hence, similar Roche lobe radii) during this phase, the donor star in the ``case B'' binary is much puffier than the one in the ``case A + case AB'' binary. The former is only partially stripped, maintaining a hydrogen-rich envelope of significant size (roughly 7 $R_\odot$, comparable to the size of the Roche lobe) and enters a phase of nuclear-timescale mass transfer in the second half of core helium burning (from $\sim$6.75 Myr onwards). This phenomenon of partial-envelope stripping and prolonged case B interactions was found by \citet{Klencki2022} to occur in massive binary models at sufficiently low metallicity, though its effect on the resulting spins has never been explored.

By contrast, the donor of the ``case A + case AB'' binary is fully stripped to a pure helium core of $\sim1\,R_\odot$ shortly after helium ignition. As the tidal torque felt by the donor is a strong function of $R_*/a$, the former star remains close to synchronization because of a much stronger torque, while in the latter case, the donor gradually loses its angular momentum due to stellar winds. Therefore, the ``case B'' binary produces a much faster-spinning black hole than the ``case A + case AB'' binary.

\begin{figure}
    \centering
    \includegraphics[width=\columnwidth]{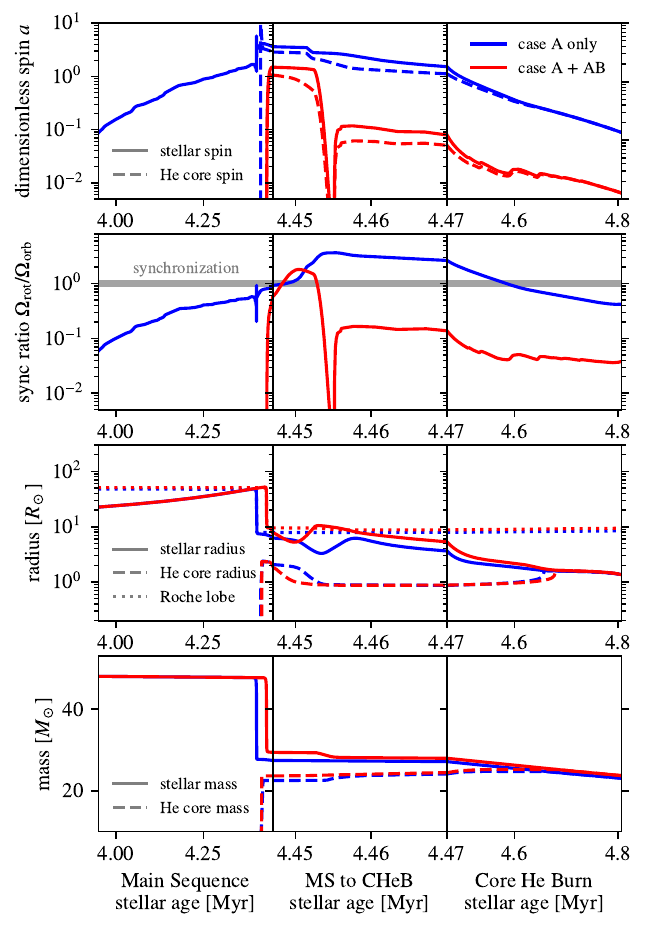}
    \caption{Comparison of the tidal spin-up history between one model that only went through case A mass transfer (blue lines) and another that went through case A + AB mass transfer (red lines). The left, middle, and right sections show different quantities zoomed in during the later stage of main-sequence, between main-sequence and core-helium ignition, and during the core-helium burning phase, respectively. We see that while the donors lose similar amounts of mass at the end, the ``case A only'' system has a higher spin than the ``case A + AB'' system. This is because the ``case A only'' system never enters a second rapid mass transfer case after the main-sequence (which happens to the ``case A + AB'' systems around 4.455 Myr), which can remove the majority of angular momentum it gains previously. Therefore, the donor remains super-synchronized when it contracts after the first (case A) mass transfer phase ends. After the main-sequence, tides are weak and the donors mainly lose angular momenta through mass loss.}
    \label{fig:compare_AB_A}
\end{figure}

Figure \ref{fig:compare_AB_A} shows a similar comparison between a ``case A only'' binary (blue lines; $M_*=50M_\odot$, $q_\mathrm{i}=0.25$, and $\log(P_\mathrm{orb,i})=1.12$) and a ``case A + case AB'' binary (red lines; $M_*=50M_\odot$, $q_\mathrm{i}=0.25$, and $\log(P_\mathrm{orb,i})=1.16$). We again chose these two models to be close to each other in the $q_\mathrm{i}-\log(P_\mathrm{orb,i})$ space, so that they evolve similarly before mass transfer occurs. The ``case A only'' model produces a much higher spin magnitude than the ``case A + case AB'' model, again confirming our findings in Section \ref{sec:results-spins}.

By comparing their evolution history, we see that the donors in both systems have a similar mass and size after mass transfer. However, the ``case A only'' system never encounters a later case B mass transfer phase that removes the majority of the accumulated angular momentum. Therefore, the donor always maintains a high rotation rate, often close to synchronization or even over-synchronized, as it contracts after the detachment from mass transfer. On the other hand, the donor in the ``case A + case AB'' binary suffers the same angular momentum loss as shown previously due to the rapid case AB mass transfer phase. The torque it feels as a fully-stripped star is weak and the resulting black hole spin becomes negligible.

We checked the evolution pathways of all our models and confirmed that the above analysis fairly represents general comparisons between different models.

\subsection{Summary of Evolution Pathways}
\label{sec:results-pathways}

\begin{figure*}
    \centering
    \includegraphics[width=\textwidth]{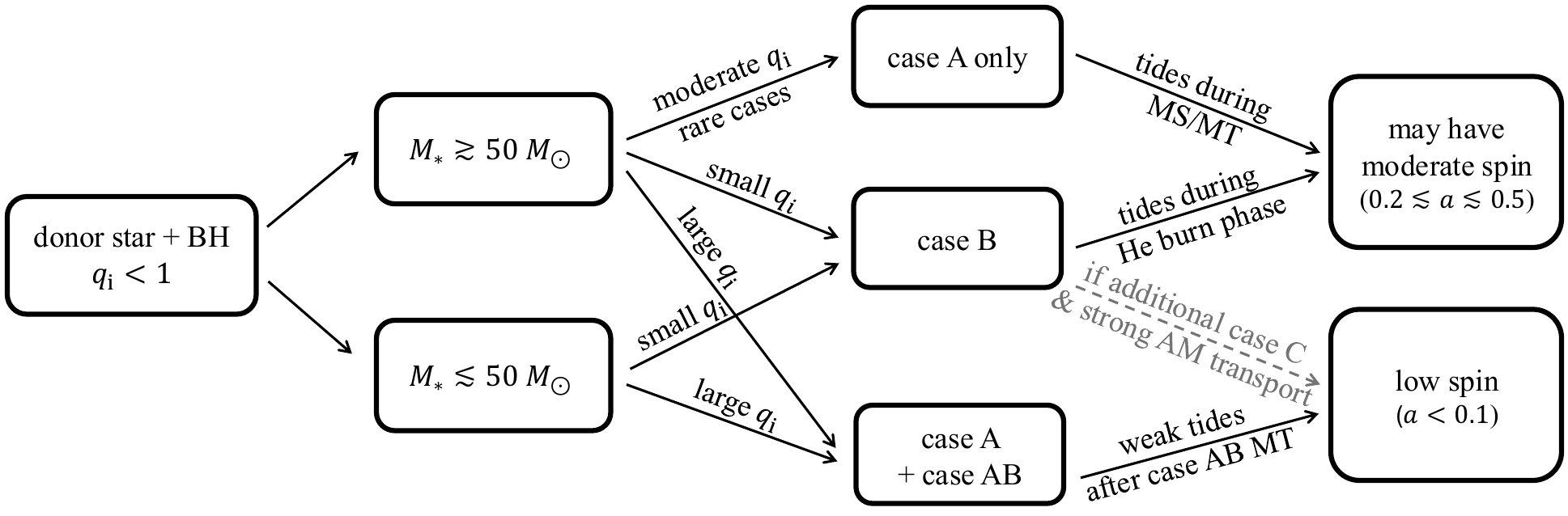}
    \caption{A flow chart summarizing our findings. Different initial mass ratios and donor masses can cause different cases of mass transfer (MT), with strong or weak tides on the donor resulting in a diverse distribution of black hole spins. If there is strong core-envelope coupling due to internal angular momentum transport before core collapse, an additional case C (also referred as case BB by some authors) mass transfer phase may bring back some of the ``case B'' systems to low spins (see discussion in Section \ref{sec:discussions_case_C}).}
    
    \label{fig:flowchart}
\end{figure*}

We see from the above comparisons that a combination of stable mass transfer cases and tides leads to a diverse spin distribution of black holes, which is in turn determined by the initial masses, mass ratios, and orbital periods of the binary. We qualitatively summarize these different pathways with the flowchart shown in Figure \ref{fig:flowchart}. 

\section{Discussion}
\label{sec:discussions}
\begin{figure*}
    \centering
    \includegraphics[width=\textwidth]{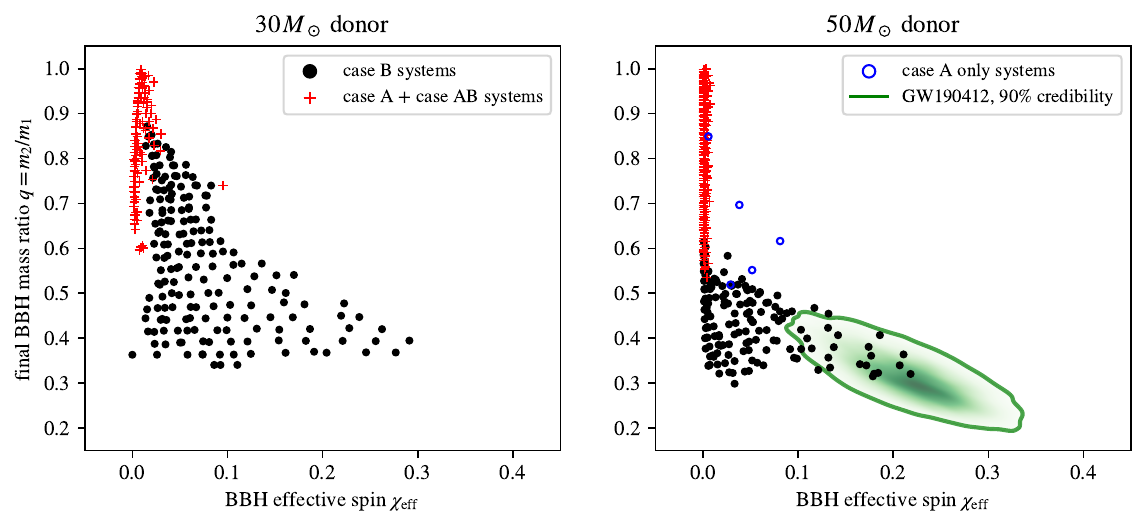}
        \caption{BBH effective spins and mass ratios derived from our results, for the $M_*=30\,M_\odot$ donor and the $M_*=50\,M_\odot$ donor. The mass ratio is defined as the ratio between the less massive black hole (secondary) to the more massive one (primary). We see that binaries that only went through case B mass transfer (black filled circles) can produce merging BBHs with $\chi_\mathrm{eff}$ between $0$ and $\sim 0.3$. In the $50\,M_\odot$-donor-mass binaries, a few systems that only went through case A mass transfer (blue empty circles) can have $\chi_\mathrm{eff}$ up to $\sim0.1$. Binaries that went through both case A and case AB mass transfer (red plus signs) rarely result in significant spins. The green region shows GW190412, a BBH event with an unequal mass ratio and significant spin of the primary, which could possibly be explained by massive binaries that went through case B mass transfer and spin-up due to tides. The area indicates 90\% credible regions for parameter estimates.}
    \label{fig:anti_corr}
\end{figure*}

\subsection{Anti-correlation between Spins and Mass-ratios}

Our results predict an anti-correlation between merging binary black hole spins and mass ratios. In Figure \ref{fig:anti_corr}, we show the effective spins and final mass ratios of binary black holes (BBHs) from our calculations. We assume that the spin is aligned with the orbit and that the first-born black hole has no spin. The final BBH mass ratio is defined as $q=m_2/m_1$, where $m_1=\max(M_\mathrm{BH},M_\mathrm{He\,core})$ is the mass of the primary and $m_2=\min(M_\mathrm{BH},M_\mathrm{He\,core})$ is the mass of the secondary.

We find that for models with a donor of $30\,M_\odot$, the resulting BBH has a primary mass between $11$ and $17$ solar-masses. Binaries that only went through case B mass transfer lead to a wide $\chi_\mathrm{eff}$ distribution between $0$ and $\sim 0.3$, and the BBH mass ratios range from $0.3$ to $0.8$. On the other hand, binaries that went through both case A and case AB mass transfer typically result in higher BBH mass ratios, up to $1$, but with negligible spins in most cases. We also see a clear anti-correlation between $q$ and $\chi_\mathrm{eff}$.

Models with the $50\,M_\odot$ donor predict similar results, with $20\,M_\odot\lesssim m_1\lesssim 28\,M_\odot$. The ``case B'' systems now predict $0<\chi_\mathrm{eff}\lesssim0.2$ and $0.3\lesssim q\lesssim 0.7$, while the ``case A + case AB'' systems produce non-spinning BBHs. A handful of systems that only went through the case A mass transfer result in $\chi_\mathrm{eff}$ up to $\sim 0.1$ at slightly higher mass ratios. We still see a clear anti-correlation between $q$ and $\chi_\mathrm{eff}$ in these models.

This anti-correlation is not na\"ively expected from the conventional tidal spin-up picture involving a binary following common-envelope ejection. In that case, the spin of the black hole binary is solely from the second born black hole, which, in the prior helium star phase, gained its angular momentum through the tidal torques exerted by the first born black hole. For a fixed semi-major axis, if the first born black hole is more massive, the tidal torque becomes stronger ($\tau\propto q^2$, see discussions in Section 4 of \citealt{Ma2023}), and the second born black hole tends to spin faster, which would predict a positive correlation between mass ratio and spins. Nevertheless, with the detailed binary mass transfer histories, we see that these two seemingly opposing results are actually consistent with each other. This is because the high-spinning systems are selected from those that only went through case B mass transfer, which occurs in wider orbits and therefore requires more unequal initial mass ratios (see Figure \ref{fig:a_q_logPi}).

There are claims of model-dependent evidence of an anti-correlation between $q$ and $\chi_\mathrm{eff}$ based on population analysis of gravitational wave data \citep{Callister2021,LIGO2023,Adamcewicz2023,LIGO2025}. It is unclear whether this anti-correlation is caused by the shifting of mean $\chi_\mathrm{eff}$ or the broadening of its distribution, though the latter is more favored in the latest GWTC-5.0 data \citep{GWTC-5.0}. Our models predict that both the mean and broadness of $\chi_\mathrm{eff}$ increase with decreasing mass-ratios, which should be tested against future data when more gravitational-wave events are detected.

There are several theoretical efforts to explain this possible anti-correlation, including BBHs formed by hierarchical mergers in dense star clusters \citep{Antonini2025b} and mergers in AGN disks \citep{MacKernan2022,Santini2023,Li2025}. Previous works on stable mass transfer in the isolated binary evolution channel, which are mostly based on rapid population synthesis, can also produce such an anti-correlation through binary mass-ratio reversal \citep{Broekgaarden2022,Zevin2022,Banerjee2023,Olejak2024,Smith2026,Chen2026,Godfrey2026,Maclean2026}.

In this picture, the high-spin BBHs form through binaries with the tightest orbits, where the donor star experiences the strongest tides to be spun-up. These binaries have very uneven mass ratios in the star-black-hole phase, as the more massive the donor is relative to the first born black hole, the more the orbit can shrink through stable mass transfer. Therefore, when the second black hole is formed, it can still be more massive than the first black hole ("mass-ratio reversal," as the first born black hole must originate from the more massive star in the initial stellar binary). The uneven mass ratio can then be passed to a small mass ratio in the BBH phase, where the more massive, second born black hole rotates rapidly.

Our results, which are based on detailed binary modeling, add a new physical recipe to this picture. We find that the high $\chi_\mathrm{eff}$ BBHs are formed primarily through binaries that went through ``case B'' mass transfer. This particular mass transfer history not only selects those uneven mass ratio binaries that would lead to mass ratio reversal, as previous works have shown, but also enhances the effect of tides by forming partially-stripped donors that experience stronger tidal torques, which can spin them up even more.

However, we note that as our models are uniformly distributed in the $q_\mathrm{i}-\log(P_\mathrm{orb,i})$ space, our results should not be compared directly with the GWTC population. Works that modeled the detailed binary evolution before the first black hole is born show that small $q_\mathrm{i}$s could be very rare (see, e.g., \citealt{Xu2025,Chen2026,Briel2026}, though the exact outcomes depend on specific assumptions of mass accretion efficiency). This could reduce the possibility of producing binaries that will go through case B mass transfer only. Future work should investigate whether the physics we propose could produce a realistic population of $q\neq1$ spinning BBHs.

\subsection{The Case of GW190412}

Among the gravitational wave merger events to date, GW190412 is unusual for its asymmetric masses and high spin \citep{LIGO2020}. The binary has a primary mass of $28.1^{+4.8}_{-4.3}\,M_\odot$ and a secondary mass of $8.8^{+1.5}_{-1.1}\,M_\odot$. The uneven mass-ratio allows for a relatively accurate measurement of its spin, with $\chi_\mathrm{eff}=0.22^{+0.08}_{-0.11}$. With a mass ratio of $0.31^{+0.12}_{-0.07}$, this $\chi_\mathrm{eff}$ is likely due to the rotation of the primary unless the secondary is maximally spinning.

We show the 90\% credible regions for parameter estimates of $q$ and $\chi_\mathrm{eff}$ of GW190412 in the right panel of Figure \ref{fig:anti_corr} (using the Phenom PHM waveform model from \citealt{LIGO2020}), along with our model predictions for the $50\,M_\odot$-donor binaries. While the black hole masses from our models are in the lower end of the measurements for GW190412, we see that the ``case B'' systems can predict a consistent mass ratio and effective spin, solely contributed by the primary. Therefore, isolated binary formation with tides and stable mass transfer could provide an alternative explanation for the origin of GW190412, other than the hierarchical merger channel (see, e.g., \citealt{Rodriguez2020}).

\subsection{Comparison to Previous Works}

\begin{figure*}
    \centering
    \includegraphics[width=\textwidth]{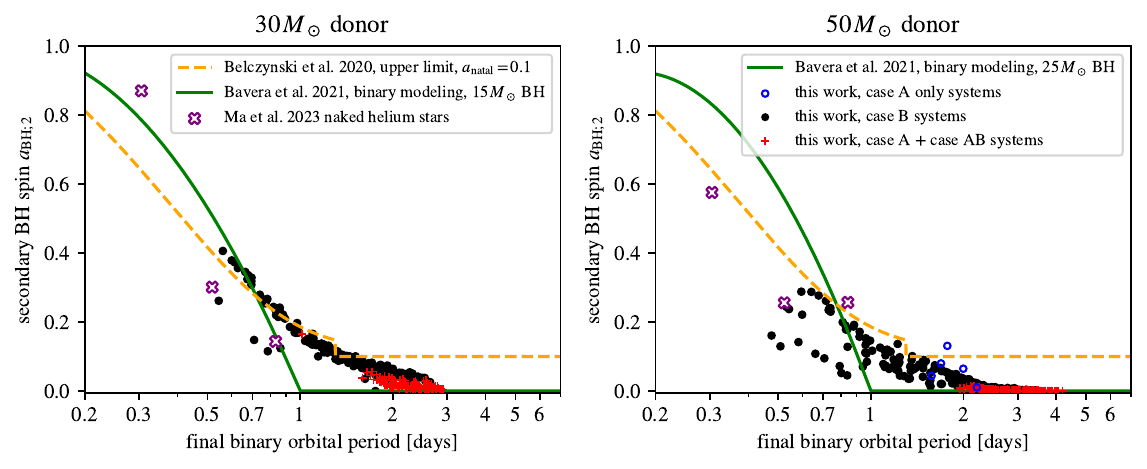}
    \caption{Second-born black hole spins compared to the results calculated from naked helium star progenitors in previous works, as a function of final binary orbital periods. The orange dashed line shows an upper limit of black hole spins by assuming tidal locking (\citealt{Belczynski2020}, Eq. 15, assuming a natal spin of 0.1 at long periods). The green line shows an analytical approximation of results from detailed binary modeling (\citealt{Bavera2021}, Eq. 1, for a $15\,M_\odot$ and $25\,M_\odot$ black hole, respectively), and the empty purple crosses show the calculations in \protect\cite{Ma2023} with the same tidal prescriptions as in this work, for helium stars at $1\%$ solar-metallicity. We see an excess of high-spin black holes at longer ($P_\mathrm{orb}\gtrsim 1\,\mathrm{day}$) orbital periods, primarily contributed by case B and case A only systems. On the other hand, fully stripped helium stars, which are only produced by ``case A + case AB'' systems in the stable mass transfer channel, can never get to $P_\mathrm{orb}\lesssim 1\,\mathrm{day}$ to get significant spins from tides. While the maximum spins are capped by the estimate by \protect\cite{Belczynski2020} as expected, our results show a broader spin distribution at given periods, especially for systems with a $50\,M_\odot$ donor.}
    \label{fig:a_Porb}
\end{figure*}

Previous works focusing on the tidal spin-up of black hole progenitor stars mostly concerned idealized helium-star-black hole binaries, where the star is fully-stripped and often assumed to be the outcome of common-envelope ejections. In Figure \ref{fig:a_Porb}, we compare our results to some of these works, showing the second-born black hole spin as a function of the final binary orbital periods.

We show a selection of previous works in addition to our results. The orange dashed line shows the analytical results from \cite{Belczynski2020}, who assumed tidal locking, which should serve as a reference for the maximum black hole spin for given orbital periods if it originates from tides. The green line shows an approximated formula from \cite{Bavera2021}, describing the conventional treatment of tides with detailed binary modeling. The purple crosses show the results from \cite{Ma2023}, who modeled the tidal spin-up using the same method as in this paper for a few $1\%$ solar-metallicity helium-star-black-hole binaries at similar masses and different orbital periods.

We see that while previous works predict that high spin black holes only exist in binaries in close orbits ($P_\mathrm{orb}\lesssim1\,\mathrm{day}$), there is a significant population of them at longer periods through the stable mass transfer channel. These black holes mostly form through ``case B'', where the progenitor star can retain a puffier hydrogen-rich envelope compared to pure-helium Wolf-Rayet stars. This allows them to feel stronger tides in the helium-burning phase and get spun-up more.

For systems with $50\,M_\odot$ donors, there is an additional population of high-spin black holes at longer periods, formed from ``case A only'' binaries. In contrast to normal Wolf-Rayet progenitors that gain their spins through tides during the helium burning phase, their progenitor stars are spun-up during the first mass-transfer phase on the main-sequence. After detachment, these stars are often super-synchronized, and the resulting spins can even be higher than those of black holes coming from tidally-locked binaries at the same orbital periods. This population was only studied as test models in full parameter-space searches before (see, e.g., \citealt{Qin2018}), and we find here a physical way to form them (see also \citealt{Xu2025}).

The stable mass transfer channel can also produce fully-stripped helium stars, as studied in previous works. However, we find that they only form from the ``case A + case AB'' systems and can never shrink to orbital periods less than $\sim 1\,\mathrm{day}$. This is consistent with the findings in \cite{Klencki2025} that there exists a fundamental limit on how close binaries can get in the stable mass transfer channel. We find that these fully stripped binaries can never get close enough to produce BBHs with significant spins.

At short periods, the maximum spins we find for BBHs is capped by the analytical formula from \cite{Belczynski2020} assuming tidal locking, which is expected. However, instead of a universal spin value at each orbital periods, our results show a broader distribution of black hole spins due to the complicated mass transfer history involved. This is most significant in the $50\,M_\odot$-donor models and is again not expected in previous works that only concerned simple helium-star-black-hole binaries.

\subsection{Hydrogen Envelopes and Case C Mass Transfer}
\label{sec:discussions_case_C}

As we saw in the previous section, a major difference in forming binary black holes through the stable mass transfer channel, compared to the conventional helium-star-black-hole binary evolution channel, is that it can produce a population of ``case B'' binaries whose donor stars still retain a residual hydrogen-rich envelope. While not significant in mass, such envelopes make the stars much puffier, allowing for stronger tides in wide orbits, which produce spinning binary black holes at longer periods.

However, the existence of residual hydrogen envelopes makes the late-stage evolution of the binary more complex. In this work, we calculated the BH spins based on the remaining angular momentum inside the helium core at the moment of central-helium depletion. From that point on, there are still tens of thousands of years left before the core collapse, mostly comprised of the phase of He-shell burning, followed by carbon burning and beyond. It was shown that helium stars with a residual H envelope are prone to expansion during He-shell burning \citep[much more compared to naked helium stars][]{Laplace2020, Klencki2022}, leading to an additional phase of ``case C'' mass transfer (\citealt{Savonije1976}, also referred as ``case BB'' mass transfer by some authors). This is relevant for the BH spins because any additional mass loss could in principle affect the angular momentum of the core.

It is challenging to accurately model how case C mass transfer affects the spin of the helium core. The degree of additional mass stripping will depend on the details of the radial expansion of helium stars, which are sensitive to their stellar winds \citep[e.g.,][]{Gilkis2019,Sander2023}, internal chemical mixing \citep[e.g.,][]{Dutta2024} including the effects of past mass accretion \citep{Xu2025}, and the numerical treatment of superadiabacity in 1D stellar codes \citep[Section 7 of][]{Jermyn2023} that affect the possible envelope inflation and stellar radii \citep[e.g.][]{Grafener2012,Romagnolo2025}.

The impact of case C mass transfer on the spin of the helium core will, in turn, depend on internal angular momentum transport that can take place on the thermal timescale during these advanced evolutionary stages. Even though our BH spin values are based on the state at core-He depletion, we evolved the binaries further until the end of central carbon burning. We find that case C mass transfer occurs in systems from the ``case B'' evolutionary pathway with $30\,M_\odot$ donors, but not in the case of $50\,M_\odot$ donors. We attribute this to the effect of stellar winds: helium stars descending from $50\,M_\odot$ donors are sufficiently strong to self-strip the residual hydrogen after case B mass transfer, preventing any further case C mass transfer\footnote{However, the wind is not strong enough to remove the angular momentum accumulated by tides. Hence, the stellar core could still result in spinning black holes. In our models, we find there is eventually a balance between the tidal torque and the wind AM loss near the end of the helium burning phase, setting an equilibrium stellar rotation rate that can be a significant fraction of the orbital frequency.}. Inspecting the binary grids from \citet{Klencki2025}, we find that the maximum donor mass that leads to case C evolution is higher at lower metallicity.

If we still assume rigid rotation of the donors after core-helium depletion (meaning that internal angular momentum could be efficiently transported on a timescale comparable to or shorter than the thermal timescale on which mass transfer occurs), we find that binaries with $30\,M_\odot$ donors always lead to negligible spins, as the ``case B'' donors will lose the majority of their core angular momentum through the additional case C mass transfer phase. On the other hand, black holes formed from the $50\,M_\odot$ donors retain their spins, as they do not undergo this additional phase of mass transfer. This indicates that more massive black holes are more likely to have high spins.

If we assume no internal angular momentum transport, the helium core will be decoupled from the stellar envelope. If the core collapses to a black hole without any further fallback from the envelope, our predictions for black hole spins remain unchanged regardless of whether case C mass transfer occurs. As the timescale for core-envelope coupling is uncertain, we mark this uncertainty as a gray arrow in the flowchart (Figure \ref{fig:flowchart}).

\subsection{Non-linear Tides}
\label{sec:nonlinear}

The mode decomposition method we are using assumes the linearity of each tidally-excited mode, which is not always guaranteed when tides are strong. A useful measure to test the linearity of modes is the maximum forced  amplitude parameter, which can be evaluated for a given mode $\xi_\alpha$ \citep{Ma2023}:
\begin{equation}
    \Phi_\alpha=\left(\frac{d\xi_{\alpha,r}}{dr}\right)_\mathrm{max}\,.
\end{equation}

When $\Phi_\alpha\ll 1$, the linearity assumption largely remains valid. As $\Phi_\alpha$ gets closer to unity, the tidally excited modes are no longer independent, and they couple weakly due to nonlinear effects. This can allow for more dissipation than linear damping \citep{Weinberg2012}. When $\Phi_\alpha>1$, the tidally excited waves reach a ``wave breaking'' amplitude as they propagate, dissipating their entire energy and angular momentum into the star \citep{Barker2011}. When this happens before the waves can reflect back near the stellar surface, it prevents them from forming standing waves (i.e., modes), and the overall tidal response is identical to the ``traveling wave limit'' originally proposed by \cite{Zahn1975}.

We check the linearity of tidally excited modes for our models. We found that when the tidal excitation is off-resonance, $\Phi_\alpha\ll 1$, and the modes largely remain linear. When a mode is resonantly excited, its amplitude is greatly enhanced and can sometimes trigger nonlinear effects. However, the exact outcome of nonlinear mode coupling is hard to predict from first principles and requires future work to investigate more.

Wave breaking almost only happens when the resonantly excited gravity waves tunnel out of the outer-most sub-surface convective regions in our stellar models. It is unclear whether this small wave propagating region actually exists in reality, as sub-surface convection is often found to reach the stellar photosphere in 3-dimensional models of hot, massive star envelopes (\citealt{Schultz2022,Schultz2023}, Ma et al. in prep.). Solving the tidal dissipation in this scenario requires 3D simulations of wave excitation and damping, which is beyond the scope of this work.

\subsection{Other Caveats}
\label{sec:caveats}

In this work, we applied the ``mode decomposition'' method, which assumes that the tidally excited modes are linearly independent when radiative damping is weak. Mathematically, this is not always guaranteed. Recently, \cite{Sun2023} developed a new method to avoid this difficulty by directly solving the tidally excited oscillations, without decomposing them into modes. \cite{Narayan2026} has been able to implement this method into binary evolutionary models. However, the computational cost remains expensive, making this implementation unsuitable for calculating a large grid of models. Future work aiming for a more accurate treatment of binary evolution and tides should investigate this new method further.

In our binary models, we parameterized the switch between stable and unstable mass transfer with a threshold mass transfer rate of $1\,M_\odot\,\mathrm{yr}^{-1}$. This is a simplification of the actual case, which depends on many more physical processes that are not included in our 1D stellar models. The exact choice of this parameter affects the lower-edge of the stable-mass-transfer ``banana shape'' that we see in Figure \ref{fig:a_q_logPi}, which matters most for the high spin systems we found. Nevertheless, changing this boundary will not change the qualitative behavior of black hole spins.

The shrinking of binary orbits during stable mass transfer sensitively depends on the orbital angular momentum carried away by the material lost from the system. For simplification, we assumed they carry the specific angular momentum near the vicinity of the accretor/at the outer Lagrangian point in our models (see detailed descriptions in Section \ref{sec:models_setup}). \cite{Scherbak2025,Scherbak2025b} recently pointed out that the mass flow from L2 can form a circumbinary disk that interacts with the ejected materials, carrying away more or less angular momentum compared to our assumptions. This can lead to tighter orbits in post-stable-mass-transfer binaries and may result in larger black hole spins. Nevertheless, for these binaries that eventually form merging BBHs, while their exact regions in the $q_\mathrm{i}-P_\mathrm{orb,i}$ parameter space may differ, they will still undergo a variety of different histories leading to qualitatively similar trends for black hole spins.

We start our donor stars from the zero-age main sequence in our models. In reality, the donor likely had a previous accretion phase due to the formation of the first-born black hole and may have a different chemical profile compared to normal single stars. This is known as ``rejuvenation'' and has been studied in the context of binary mass transfer \citep{Renzo2023}. Future work should investigate how this will affect stable mass transfer and black hole spins.

We calculated the tidal evolution based on post-processing of pre-calculated binary orbits. This creates an inconsistency, as the action of tides on orbits cannot be updated in real time. \cite{Detmers2008} suggests that when tides are strong, massive stellar winds can effectively carry away a lot of orbital angular momentum, as the orbit is always synchronized with the rotation of the donor. This may lead to the merger of the binary before the second black hole can form, preventing the binary black hole from forming. Future work should also look into this possibility if a consistent treatment of tides and binary orbital evolution can be developed.

\section{Conclusions}
\label{sec:conclusions}
In recent years, there have been numerous works showing that stable mass transfer (SMT) can be an important channel for forming merging binary black holes (BBHs) that are detected as gravitational-wave signals \citep{vandenHeuvel2017,Marchant2021,Klencki2021,Olejak2021,Gallegos-Garcia2021,vanSon2022,Picco2024,Xu2025}. In this work, we examine the spins of binary black holes formed through this channel. We established binary evolution models with a self-consistent treatment of mass transfer and calculated the tidal spin up of the donor star through post-processing.

We found that the stable mass transfer channel can form BBHs with non-negligible spins, but the exact outcome depends on the binary mass transfer history. For systems with a $30\,M_\odot$ donor, they may go through a case A mass transfer phase followed by a case AB mass transfer phase that fully strips the donor star. The final black hole spin when the donor collapses is small. If the binary is initially in a wider orbit with a less massive secondary, it could trigger a case B mass transfer phase that does not fully strip the donor, and the resulting black hole may have moderate spin ($0.2\lesssim a\lesssim 0.5$) even when the binary is in longer-period orbits. For systems with more massive donors ($M_\mathrm{donor}\gtrsim50\,M_\odot$), they may occasionally have only one case A mass transfer phase, allowing the donor to be synchronized during mass transfer. This is an additional channel that leads to moderate spins (up to $a\sim 0.2$ in our models), different from the conventional tidal spin-up picture in which stars get spun-up during the core-helium burning phase.

We predict an anti-correlation between the binary black hole mass ratios and spins, consistent with some limited evidence found in gravitational wave data. This is because systems with high spins primarily form from ``case B'' or ``case A only'' systems that must start with a smaller companion to undergo stable mass transfer and form merging black holes. Our ``case B'' models have the potential to explain GW190412, a moderately-spinning ($\chi_\mathrm{eff}=0.22$) binary black hole with an uneven mass ratio of $0.31$ that cannot be explained through the conventional tidal spin up channel. Overall, stable mass transfer can produce a broad range of black hole spins between 0 and $\sim 0.5$, consistent with the evidence for a broad spin distribution in gravitational-wave data.

Our work has some caveats. If the donor star does not lose its entire envelope at the end of the helium burning phase, it may enter an additional case C mass transfer phase during carbon burning before core collapse. This primarily happens to our $30\,M_\odot$-donor models. If internal angular momentum transport efficiently couples the stellar core and envelope in this phase, the core could lose the majority of its angular momentum, resulting in low spins. However, the efficiency of angular momentum transport in this phase is highly uncertain and should be investigated in future works.

We discussed other uncertainties in our work, including limitations of the mode decomposition method we used, limitations on the linear wave approximation, uncertainties in some of our parameter choices, and additional physics that is not included in our models, such as L2 mass transfer. We expect our overall qualitative conclusions will remain largely unaffected despite these uncertainties.

\section*{Acknowledgements}

We thank Pablo Marchant, Monica Gallegos-Garcia, Ruggero Valli, Matias Zaldarriaga, Norbert Langer, Aleksandra Olejak, Tejaswi Venumadhav, Matthieu Renzo, Sterl Phinney, Vicky Kalogera, Ilya Mandel, Jim Fuller, Selma de Mink, and Maya Fishbach for useful discussions. LM is supported by the Lyman Spitzer, Jr. Postdoctoral Fellowship of Princeton University and the Gordon and Betty Moore Postdoctoral Award of the Kavli Institute for Theoretical Physics (KITP). LvS is supported by VI.Veni.242.115, Grant ID \url{https://doi.org/10.61686/XVIAV86753}. The collaboration between the authors was supported in part by the Munich Institute for Astro-, Particle and BioPhysics (MIAPbP), which is funded by the Deutsche Forschungsgemeinschaft (DFG, German Research Foundation) under Germany's Excellence Strategy - EXC-2094 - 390783311, and in part by grant NSF PHY-2309135 to the Kavli Institute for Theoretical Physics.  This work also benefited from interactions supported by the Gordon and Betty Moore Foundation through grant GBMF5076. 

\section*{Data Availability}

The data and source code supporting the plots within this article are available on reasonable request to the corresponding author.



\bibliographystyle{mnras}
\bibliography{bibliography}


\appendix

\section{Integration Time Steps}
\label{app:integration_timesteps}

When integrating the spin-orbit evolution, we select the integration time steps adaptively to achieve a smooth variation of tidal torques for numerical convergence. From Equation \ref{eq:torque_mode}, the torque depends sensitively on the stellar structure (hence, the eigenmode solutions from it) and how close the orbit is to the resonance of certain eigenmodes.

We integrate the evolution from $t_\mathrm{start}$ to $t_\mathrm{end}$ and select a global upper limit of
\begin{equation}
    \Delta t_\mathrm{evol}=10^{-3}\,(t_\mathrm{end}-t_\mathrm{start})\,,
\end{equation}
which ensures that the full integration has at least approximately $10^3$ steps. We additionally restrict he timestep to reach the next available MESA snapshot, preventing the integration from skipping changes in the interpolated stellar structure.

$\tau_\mathrm{tide}$ has a zero point when the binary reaches synchronization and the tidal forcing frequency $\omega_\mathrm{f}=0$. This imposes a further constraint based on $\omega_\mathrm{f}$ and its instantaneous derivative. Away from synchronization, defined by $|\Omega_\mathrm{orb}-\Omega_\mathrm{spin}|>10^{-3}\Omega_\mathrm{orb}$, we limit the timestep to 1\% of the estimated time required for the forcing frequency to reach zero:
\begin{equation}
    \Delta t_\mathrm{sync}=10^{-2}\left|\frac{\omega_\mathrm{f}}{\dot{\omega_\mathrm{f}}}\right|\,.
\end{equation}
This prevents a single integration step from overshooting the synchronized state. Once the system lies within the synchronization tolerance, we relax the restriction to
\begin{equation}
    \Delta t_\mathrm{sync}=2.01\left|\frac{\omega_\mathrm{f}}{\dot{\omega_\mathrm{f}}}\right|\,,
\end{equation}
so that synchronization can be reached in a finite number of steps. This restriction is not applied when $\dot{\omega}_\mathrm{f}=0$.

When calculating torques using the interpolated eigenmode properties, we introduce an additional resonance-resolving timestep, using the mode whose eigen-frequency $\omega_\alpha$ is closest to $\omega_\mathrm{f}$. Near resonance, defined as $|\omega_\mathrm{f}-\omega_\alpha|<|\gamma_\alpha|$, we limit the timestep by:
\begin{equation}
\Delta t_\mathrm{mode}=0.1\,\min\left(\left|\frac{\gamma_\alpha}{\dot{\omega}_\mathrm{f}}\right|,\left|\frac{\omega_\alpha}{\dot{\omega}_\alpha}\right|\right)\,.
\end{equation}
These conditions limit the forcing-frequency change to one tenth of the resonance width and resolve the evolution of the mode frequency due to changes in stellar structure. Away from resonance, we replace the constraint by
\begin{equation}
\Delta t_\mathrm{mode}=0.1\,\min\left(\left|\frac{\omega_\mathrm{f}-\omega_\alpha}{\dot{\omega}_\mathrm{f}}\right|,\left|\frac{\omega_\alpha}{\dot{\omega}_\alpha}\right|\right)\,,
\end{equation}
so that the calculation cannot traverse the frequency separation to the nearest resonance in a single step.

The adopted timestep is therefore
\begin{equation}
    \Delta t=f_{\Delta t}\max\left[\min(\Delta t_\mathrm{evol},\Delta t_\mathrm{sync}, \Delta t_\mathrm{mode}), 1\,\mathrm{day}\right]\,,
\end{equation}
where $f_{\Delta t}$ is a timestep scaling factor. Thus, the most restrictive physical or numerical criterion is normally used, subject to a one-day timestep floor. We did our integration for $f_{\Delta t} = 1.0$ and $0.5$ and confirmed the numerical convergence.


\bsp	
\label{lastpage}
\end{document}